\documentclass[aps,prd,final,letterpaper, nofootinbib, superscriptaddress]{revtex4}

\usepackage[utf8]{inputenc}
\usepackage{amsmath,amsfonts,amssymb,amsthm,graphicx,textcomp}
\usepackage{hyperref}

\def\braket#1{\mathinner{\langle{#1}\rangle}}
\newcommand{\eq}[1]{\begin{align}#1\end{align}}

\def\du{\ensuremath{\mathrm{d}}}

\newcommand{\unrho}{{\widetilde{\rho}}}

\newcommand{\eV}{\text{\;eV}}
\newcommand{\keV}{\text{\;keV}}
\newcommand{\MeV}{\text{\;MeV}}
\newcommand{\GeV}{\text{\;GeV}}
\newcommand{\TeV}{\text{\;TeV}}

\newcommand{\Fig}[1]{Fig.~\ref{#1}}
\newcommand{\Eq}[1]{Eq.~(\ref{#1})}
\newcommand{\Sec}[1]{Sec.~\ref{#1}}
\newcommand{\App}[1]{Appendix~\ref{#1}}

\begin{document}

%opening
\title{The Stochastic Axion Scenario}

\author{Peter~W.~Graham}
%\email{pwgraham@stanford.edu}
\affiliation{Stanford Institute for Theoretical Physics, Department of Physics, Stanford University, Stanford, CA 94305}

\author{Adam Scherlis}
\affiliation{Stanford Institute for Theoretical Physics, Department of Physics, Stanford University, Stanford, CA 94305}

\begin{abstract}
For the minimal QCD axion model it is generally believed that overproduction of dark matter constrains the axion mass to be above a certain threshold, or at least that the initial misalignment angle must be tuned if the mass is below that threshold.  We demonstrate that this is incorrect.  During inflation the axion tends toward an equilibrium, assuming the Hubble scale is low and inflation lasts sufficiently long.  This means the minimal QCD axion can naturally give the observed dark matter abundance in the entire lower part of the mass range, down to masses $\sim 10^{-12}\eV$ (or $f_a$ up to almost the Planck scale).  The axion abundance is generated by quantum fluctuations of the field during inflation.  This mechanism generates cold dark matter with negligible isocurvature perturbations.  In addition to the QCD axion, this mechanism can also generate a cosmological abundance of axion-like particles and other light fields.

%During inflation, light scalar fields gradually approach an equilibrium energy density of order $H_I^4$. If $H_I<\Lambda_{QCD}$, this effect determines the distribution of initial misalignment angles for the ``anthropic'' QCD axion, solving the overclosure problem without anthropics and providing a new production mechanism for axion dark matter. The resulting dark matter is cold and does not violate isocurvature constraints. This opens up a new swath of parameter space for QCD axion dark matter, allowing for natural dark matter densities at arbitrarily high $f_a$ for a small range of $H_I$. For $H_I$ significantly below this range, the anthropic axion is impossible.
\end{abstract}

\maketitle

\tableofcontents

\section{Introduction}
The identity of dark matter is one of the major outstanding questions in physics. The mass of the dark matter particle is enormously unconstrained, ranging from $10^{-22}\eV$ for fuzzy dark matter models
to the Planck scale $\approx 10^{27}\eV$, or even higher for dark matter composed of primordial black holes.

The QCD axion is a popular and well-motivated candidate for light ($m_a \ll \eV$) dark matter. In the  ``post-inflationary'' scenario for axion dark matter, the population of axions is produced after inflation, when Peccei-Quinn symmetry breaks spontaneously. This scenario predicts a unique value for the axion mass in the {\textmu}eV range (or axion decay constant $f_a \sim 10^{12}\GeV$). However, if PQ symmetry is already broken during inflation, the axion mass can be in a broad range roughly $10^{-12}-10^{-4}\eV$ or even beyond and still be dark matter.
With a relatively short duration for inflation, the axion abundance is set by the initial average value of the axion field, $f_a\theta_0$ (where $\theta_0$ is the initial misalignment angle). In order for the axion to be much lighter than the post-inflationary value, $\theta_0\ll 1$ is required. This has been treated in the past as a fine-tuned initial condition, or even as a sign of anthropic selection (see e.g.~\cite{Linde:1987bx,Linde:1991km,Tegmark:2005dy,Hertzberg:2008wr}).  However, there are many  models of early universe cosmology in which such a low-mass (high-$f_a$) QCD axion can in fact naturally have the correct dark matter abundance without tuning or anthropics (see e.g.~\cite{Agrawal:2017eqm, Nomura:2015xil,  Dine:1982ah,Steinhardt:1983ia,Lazarides:1990xp,Kawasaki:1995vt,Dvali:1995ce,Choi:1996fs,Banks:1996ea,Banks:2002sd}).   These models generically have new particles coupling to the axion or the Standard Model.

The question of whether axion dark matter can naturally have a light mass or high $f_a$ is an important one.  The axion is an extremely well-motivated dark matter candidate and significant effort is being expended to search for it (see e.g.~\cite{Du:2018uak, Zhong:2018rsr, Anastassopoulos:2017ftl, TheMADMAXWorkingGroup:2016hpc, Geraci:2017bmq, Graham:2017ivz, Ehret:2009sq, Graham:2015ouw} and many others).  The axion mass is critical to determining the type of experimental search used to detect it.  Any guidance from theory on the axion mass is therefore important.  Lighter axions arise from PQ-breaking scales around fundamental scales such as the grand unified theory (GUT), string, or Planck scale and are therefore theoretically very well motivated (see e.g.~\cite{Svrcek:2006yi, Arvanitaki:2009fg,Cicoli:2012sz,Ernst:2018bib,Reig:2018ocz}).  There are in fact several experiments aiming to detect axions with low masses.  In particular CASPEr \cite{Budker:2013hfa, Graham:2013gfa, Garcon:2017ixh, JacksonKimball:2017elr} aims to push down to the QCD axion at the lowest masses, $f_a$ around the GUT to Planck scales.  Additionally, electromagnetic LC-circuit experiments, e.g. LC Circuit \cite{sikivie2014proposal}, DM Radio \cite{Chaudhuri:2014dla, Silva-Feaver:2016qhh}, or ABRACADABRA \cite{Kahn:2016aff}, aim to push to QCD axion masses somewhat below cavity searches in the ``post-inflationary'' region.  Hopefully the combination of all axion experiments will be able to cover the entire allowed QCD axion mass range.

\section{Summary}

In this paper we observe that even the simplest QCD axion model can have the correct dark matter abundance at low axion mass, new particles are not required.  All that is necessary is a long period of low-scale inflation.  During  a long enough period of inflation, the axion is naturally driven to an equilibrium distribution of field values.  This equilibrium can produce the correct dark matter abundance under one constraint on the  parameters: the Hubble scale of inflation $H_I$ and the mass of the axion $m_a$.  We show our results for this region in \Fig{bigplot}.

For long durations of inflation, if the Hubble scale $H_I$ is lower than the QCD scale, $\theta$ relaxes to a small value\footnote{Note that such a period of long, low-scale inflation is also used by many relaxion \cite{Graham:2015cka} solutions to the hierarchy problem.}~\cite{Dimopoulos:1988pw}.
The axion abundance is naturally suppressed because the axion is classically driven to the minimum of its potential during inflation.  However the axion abundance cannot go to zero because of quantum fluctuations.  A nonzero abundance is generated by quantum fluctuations of the axion field during inflation.  The competition between the classical force driving the axion field towards zero and the stochastic quantum fluctuations which can drive the axion field up its potential results in an equilibrium.  The axion field is naturally driven towards this equilibrium during inflation.  Roughly the quantum fluctuations of the axion can be viewed as arising from the de Sitter temperature during inflation.  The axion abundance can be estimated from the ``thermal'' equilibrium value of the axion field during inflation, as discussed in greater detail below.
%The Hubble scale during inflation can be viewed as a temperature, the de Sitter temperature.  The equilibrium a. 
This makes the initial $\theta$ after inflation naturally small but nonzero.
%In this paper, we observe that quantum fluctuations during a long period of inflation allow $\theta$ to be naturally small but nonzero.
And so, in particular, if $H_I$ is in the correct range, the axion naturally gives the measured dark matter abundance for axion masses as small as $\sim 10^{-12}\eV$. 
\section{Dynamics of Axions During Inflation}

Inflationary fluctuations of spectator fields are usually considered in the context of modes whose wavelengths are within the Hubble horizon today -- that is, they are visible in the power spectrum of a field. However, if inflation lasts longer than 60 e-folds, it also produces modes which never re-enter the horizon. These modes are observable because they contribute to the average value of any spectator field over our Hubble patch (see e.g.~\cite{Starobinsky:1994bd,Enqvist:2013kaa,Enqvist:2014zqa,Nurmi:2015ema,Kainulainen:2016vzv,Hardwick:2017fjo,Hardwick:2017qcw,Hardwick:2018sck}). In particular, for the QCD axion, this average value (in the early universe) is the misalignment angle responsible for post-inflationary axion dark matter. If inflation lasts long enough, the inflationary contribution dominates and the Hubble scale of inflation determines the misalignment angle and axion abundance. The distribution of misalignment angles for the ``unrolled'' axion field is multimodal and continues to evolve indefinitely~\cite{Palma:2017lww}.

%If we treat the axion field range as compact, identifying $\theta$ with $\theta+2\pi n$, the distribution approaches an equilibrium after a characteristic relaxation time. However, the resulting cosmic variance of the dark matter abundance is significant. For $H$ below the QCD scale, the distribution of misalignment angles across many Hubble patches is Gaussian with average energy density of order $H_I^4$ at reheating. For $H$ above the QCD scale, it is uniform, providing a concrete justification for assuming a uniform prior for the ``anthropic axion''~\cite{Hertzberg:2008wr}.\\

During inflation, the average values of scalar fields fluctuate around the minimum of the field's potential $V$. For light fields, which are overdamped by Hubble friction, these fluctuations can be quite large if inflation lasts long enough for them to accumulate. A free scalar $(V=\frac12m^2\phi^2)$ will typically end up with energy density of order $V\approx T_{dS}^4$, where $T_{dS}=H_I/2\pi$ is the de Sitter temperature; the field will therefore be displaced by $\phi\approx H_I^2/m$ from its minimum.

As inflation progresses, modes of the field are stretched until their wavelengths are longer than the Hubble radius, after which they are mostly ``frozen''. Each mode has amplitude of order $H_I$, and gives a random ``kick'' of this size to the average field value as it crosses the horizon, producing random-walk behavior for the field value. The field also slow-rolls down the potential towards its minimum (if it is massive with $m<H_I$); without the quantum fluctuations, this would relax the field to a very small value. These two opposing tendencies are in equilibrium when the average field value is of order $H_I^2/m$; while the average motion of the field value is always towards zero, the random fluctuations prevent it from settling exactly there.

The QCD axion is massive as long as $T\lesssim \Lambda_{QCD}$, with mass $m_a\approx \Lambda_{QCD}^2/f_a$ and CP-violating angle $\theta=\phi/f_a$. Therefore, if $H_I$ is slightly below $\Lambda_{QCD}$, inflationary fluctuations will produce a misalignment angle of order $\theta\approx H_I^2/m_af_a\approx H_I^2/\Lambda_{QCD}^2$. The misalignment required for axion dark matter is between $10^{-4}$ and $O(1)$ depending on $f_a$. QCD axion dark matter can easily be obtained from these fluctuations for any $f_a$ between $10^{11}\GeV$ and the Planck scale, depending on $H_I$. Because the fluctuations are homogeneous over super-Hubble scales, the cosmic variance in $\theta$ is high and the relationship between $f_a$ and $H_I$ is approximate.
% 
% Different Hubble patches have different values of $\theta$ due to this variance. The distribution of misalignment angles over many patches approaches an equilibrium distribution with a characteristic relaxation time of $H_I^2/m^2$ e-folds. If inflation does not last this long, the misalignment angle will be set by initial conditions.

A similar mechanism to ours has been considered for vector dark matter~\cite{Graham:2015rva}; however, for vectors, sub-horizon modes (generated in the last 60 or less e-folds) dominate, the long-wavelength modes considered in this paper are negligible, and the vector dark matter abundance has low cosmic variance.

%As noted several decades ago, MeV-scale inflation provides a solution to the overproduction problem for an axion with an initially $O(1)$ value for $\theta$~\cite{Dimopoulos:1988pw}.\\

\subsection{Setup}
Consider the average value of a scalar field (with $m\ll H_I$) over one Hubble patch. Throughout this paper, we will refer to this simply as $\phi$.
This value is approximately given by the sum of long-wavelength modes with $k\ll aH_I$, which are homogeneous on the scale of a Hubble patch; shorter wavelengths $k\gg aH_I$ are averaged over many periods and contribute much less to $\phi$. As the scalar field evolves during inflation, the $k\ll aH_I$ modes are ``frozen'', but decay slowly towards zero; equivalently, $\phi$ slow-rolls towards the minimum of its potential, overdamped by Hubble friction. Meanwhile, new modes with $k\approx aH_I$ ``leave the horizon'' and become part of $\phi$, with each order of magnitude in $k$ contributing an amplitude of order $H_I$. The phase is random, so these new modes produce a random-walk behavior for $\phi$.

Instead of focusing on the stochastically-evolving value $\phi$ in a specific Hubble patch, we will track the distribution $\rho(\phi,t)$ which describes the frequency of different values of $\phi$ across many patches. The fraction of patches with field values in a small interval between $\phi$ and $\phi+\du\phi$ is given by $\rho(\phi)\du\phi$, and $\int\rho\du\phi=1$. This distribution evolves deterministically. The potential for $\phi$ tends to concentrate the distribution near its minimum, while the random contributions cause $\rho$ to diffuse and even out. These opposing effects determine the typical width of $\rho$ when it reaches equilibrium.

We assume that the potential $V(\phi)$ is negligible compared to $V_I$, the energy density due to the inflaton; this assumption holds up for the QCD axion as long as $H_I\gg 10^{-12}\eV$. For more discussion of backreaction effects, see \Sec{backrx} and \App{fokk}.

\subsection{Massive Free Scalar}
As a simple example, consider a free scalar field $V=\frac12m^2\phi^2$ and Gaussian initial condition 
\begin{align}
\rho(\phi,0)\propto \exp({-(\phi-\mu)^2/2\sigma^2}).
\end{align}
In this case, the distribution remains Gaussian for all time.

If there were no fluctuations, every point in the distribution would independently evolve according to the slow-roll equation,
\begin{align}
\dot\phi &= \frac{V'(\phi)}{3H_I}=-\frac{m^2}{3H_I}\phi
\end{align}
This has the solution $\phi(t)\propto\exp(-(m^2/3H_I)t)$. As each point $\phi$ follows this trajectory, the distribution concentrates near zero. More precisely, the distribution $\rho(\phi,t)$ is uniformly rescaled by a factor of $\exp(-(m^2/3H_I)t)$ horizontally and $\exp((m^2/3H_I)t)$ vertically.
Therefore, the mean and standard deviation of the Gaussian would follow the same evolution,
\begin{align}
\left.\dot\mu\right|_{\text{no diffusion}} &= -\frac{m^2}{3H_I}\mu\\
\left.\dot\sigma\right|_{\text{no diffusion}}&= -\frac{m^2}{3H_I}\sigma
\end{align}
Instead of the standard deviation, it will be more useful to work in terms of the variance $\sigma^2$:
\begin{align}
\left.\dot{(\sigma^2)}\right|_{\text{no diffusion}}=2\sigma\dot\sigma=-\frac{2m^2}{3H_I}\sigma^2
\end{align}
Now consider the case of diffusion with the potential $V$ switched off. The Gaussian spreads, $\sigma\sim\sqrt t$, so the variance increases linearly:
\begin{align}
\left.\dot{\mu}\right|_{\text{no rolling}}&=0\\
\left.\dot{(\sigma^2)}\right|_{\text{no rolling}}&=\frac{H_I^3}{4\pi^2}
\end{align}
With both effects present, we have
\begin{align}
\dot\mu &= -\frac{m^2}{3H_I}\mu\\
\dot{(\sigma^2)} &= \frac{H_I^3}{4\pi^2}-\frac{2m^2}{3H_I}\sigma^2
\end{align}
The first term in $\dot{(\sigma^2)}$ is the diffusion (random walking) of the field, which tends to increase variance linearly; the second term is due to the quadratic potential squeezing the field closer to its minimum. These two effects cancel out at an equilibrium value of $\sigma^2$ given by 
\begin{align}
\sigma_f^2= \frac{3H_I^4}{8\pi^2m^2}
\end{align}
Note that this corresponds to the general formula for the equilibrium distribution given in \Eq{eq:equil}. In terms of $\Delta=\sigma^2-\sigma_f^2$, we have
\begin{align}
\dot\Delta &= -\frac{2m^2}{3H_I}\Delta
\end{align}
Therefore, the mean and variance of the distribution both approach their equilibrium values exponentially at a rate of order $m^2/3H_I$. If the field begins at a single initial value $M$ (Gaussian with $\mu(0)=M,\sigma(0)=0$), it will closely approximate the equilibrium distribution after a small multiple of 
\begin{align}
\frac {H_I}{m^2}\log(M/\sigma_f).
\end{align}
Note that a time of order $H_I/m^2$, during inflation, is $N=H_I^2/m^2$ e-folds.

Because the Fokker-Planck equation for this system is linear, any initial distribution restricted to $[-M,M]$ will reach equilibrium on the same timescale. More rigorously, we can find the full set of quasinormal modes of the Fokker-Planck equation and show that the slowest decay rates are of order $m^2/3H_I$; see \App{fokk}.

\subsection{Massless Compact Scalar}
\label{compact}
Now consider a scalar field $\phi=f\theta$ with a compact field range, so that $\phi=0$ and $\phi=2\pi f$ are identified. In this case, there is no potential so $\rho$ diffuses into a uniform distribution.

Starting from a Dirac delta, the variance of the distribution for the field, $f\theta$, will increase at a rate of $H_I^3/4\pi^2$. The variance of $\theta$ will therefore increase at a rate of $H_I^3/4\pi^2f^2$. The distribution will become uniform once this variance is $\gtrsim1$, which will take a time of order $4\pi^2f^2/H_I^3$, so that $N\gtrsim 4\pi^2f^2/H_I^2$.

A more precise calculation (\App{fokk}) gives a relaxation time of $8\pi^2f^2/H_I^2$ e-folds for this case. The quasinormal modes are sinusoidal in this case.

\subsection{General Scalar}
The cases above are easy to work with and are good approximations to the QCD axion potential for $H_I\ll\Lambda_{QCD}$ and $H_I\gg\Lambda_{QCD}$ respectively, as we will see in \Sec{QCD}.

More generally, any initial probability distribution for the value of a scalar field will asymptotically approach an equilibrium distribution\cite{Starobinsky:1994bd} (see \App{fokk}) which looks like a Boltzmann distribution with temperature of order $H_I$:
\begin{align}\label{eq:equil}
\rho_f(\phi,t) \propto \exp\left(-\frac{8\pi^2V(\phi)}{3H_I^4}\right)
\end{align}
This does not, however, mean that the field is thermalized; rather, it has a misalignment angle which gives its homogeneous mode an energy density of order $H_I^4$ (unless the potential is very flat, as in the compact case above). This energy behaves as dark energy while $H_I\gg m$ and becomes a condensate of cold matter at $H_I\lesssim m$. As we will show, this matter quickly becomes nonrelativistic, contrary to what would be expected from a bath of radiation produced at temperature $H_I$.

The distribution of field values is described by a Fokker-Planck equation (\App{fokk}) with a diffusion term (for the kicks) and a classical force term (for the classical slow-rolling). These two opposing tendencies bring the distribution to the above equilibrium, over some characteristic relaxation time. This relaxation time is the time required for inflationary super-horizon modes to dominate over whatever long-wavelength modes existed at the beginning of inflation; in other words, the timescale at which sensitivity to initial conditions goes away. Note that the variance of the distribution is entirely cosmic variance: each patch takes on one average value by definition, so we can only observe one sample from this distribution.

\section{Sub-Horizon Modes and Isocurvature Bounds}
The modes smaller than the horizon (after inflation) produce inhomogeneities in the field, which are observable as isocurvature perturbations. These modes are produced during the last 60 or less e-folds, so the mechanism we consider does not affect them.  As $H_I$ becomes smaller, these perturbations become negligible relative to the total density; for the QCD axion, they are below current observational bounds for $H_I\lesssim\TeV$ and $f_a\gtrsim10^{10}\GeV$. In particular, there is negligible isocurvature for $H_I\lesssim\Lambda_{QCD}$, where the axion mass is nonzero. For $f_a\lesssim10^{10}\GeV$, if the QCD axion makes up most of dark matter, the value of $\theta$ approaches $\pi$ and anharmonic effects increase the isocurvature perturbations~\cite{Kobayashi:2013nva,diCortona:2015ldu}.

For low $H_I$, the dark matter produced by our mechanism during long periods of inflation is extremely cold; this is due to the fact that all of the modes produced significantly prior to the last 60 e-folds have unobservably small gradients. It is fairly straightforward to calculate the velocity of dark-matter axions at matter-radiation equality: the total energy density of dark matter is comparable to radiation, so it is parametrically $T^4$ (there are only a few relativistic degrees of freedom at this point). Now, consider the kinetic energy density of all modes produced during an e-fold of inflation. When these modes re-enter, they have kinetic energy density $(k/a)^2\phi_k^2\approx H^2H_I^2\approx (H_I/M_P)^2 T^4$. If they are relativistic $(k/a>m)$, they will redshift as $a^{-4}\sim H^2$ (during the radiation-dominated era), so this formula will hold for {\it all} relativistic modes at any point until matter-radiation equality. Therefore, the kinetic energy density of the axion field is parametrically $(H_I/M_P)^2 T^4$, disregarding a logarithm. The velocity of the axions is then
\eq{v^2 \approx p^2/m^2 \approx H_I^2/M_P^2\\
v \approx H_I/M_P}
The value of $H_I/M_P$ ranges from $10^{-20}$ to $10^{-8}$ for our parameter space.

At much earlier times, when $H>m_a$, the kinetic energy density of axions is still $(H_I/M_P)^2 T^4$ but the total density is $H_I^4$. This gives $v\approx H/H_I\approx (T/T_{\text{rh}})^2$, where $T_{\text{rh}}$ is the reheating temperature. (This is as expected; kinetic energy of relativistic modes dilutes as $a^{-4}$ but total energy density is constant at this stage.)  Therefore, the axion is nonrelativistic immediately after reheating, and becomes cold very quickly.

%After the axion begins to oscillate $(H<m_a)$, the relativistic modes continue to dilute as $a^{-4}$ but the total energy now dilutes as $a^{-3}$, so $v^2\sim a^{-1}$.
%above line is only true if it oscillates after confinement, otherwise more complicated

\section{Results}
\subsection{QCD Axion}
\label{QCD}
The QCD axion has a mass which depends on temperature, which we take to be $T_{dS}=H_I/2\pi$. However, this mass has a constant value of $m_a\sim\Lambda_{QCD}^2/f_a$ for $H_I\lesssim\Lambda_{QCD}$
, where $\Lambda_{QCD}$ is the QCD scale. To be more precise, however, we should disambiguate two definitions of the QCD scale. One is based on the maximum height of the axion potential, which is $\chi(0)\approx (75\MeV)^4$ \cite{Borsanyi:2016ksw}; this is the topological susceptibility of QCD, $\chi(T)$, at zero temperature. The other definition is $T_c\approx 130\MeV$; this is the temperature below which the axion mass is constant \cite{Borsanyi:2016ksw}, near the QCD phase transition.

Incorporating this distinction, we find that $m_a=\sqrt{\chi(0)}/f_a\approx (75\MeV)^2/f_a$ when $H_I\lesssim 2\pi T_c\approx 800\MeV$. This order-of-magnitude separation in scales will be important.

\begin{figure}[h!]
	\centering
	  \makebox[\textwidth][c]{\includegraphics[width=.45\textwidth]{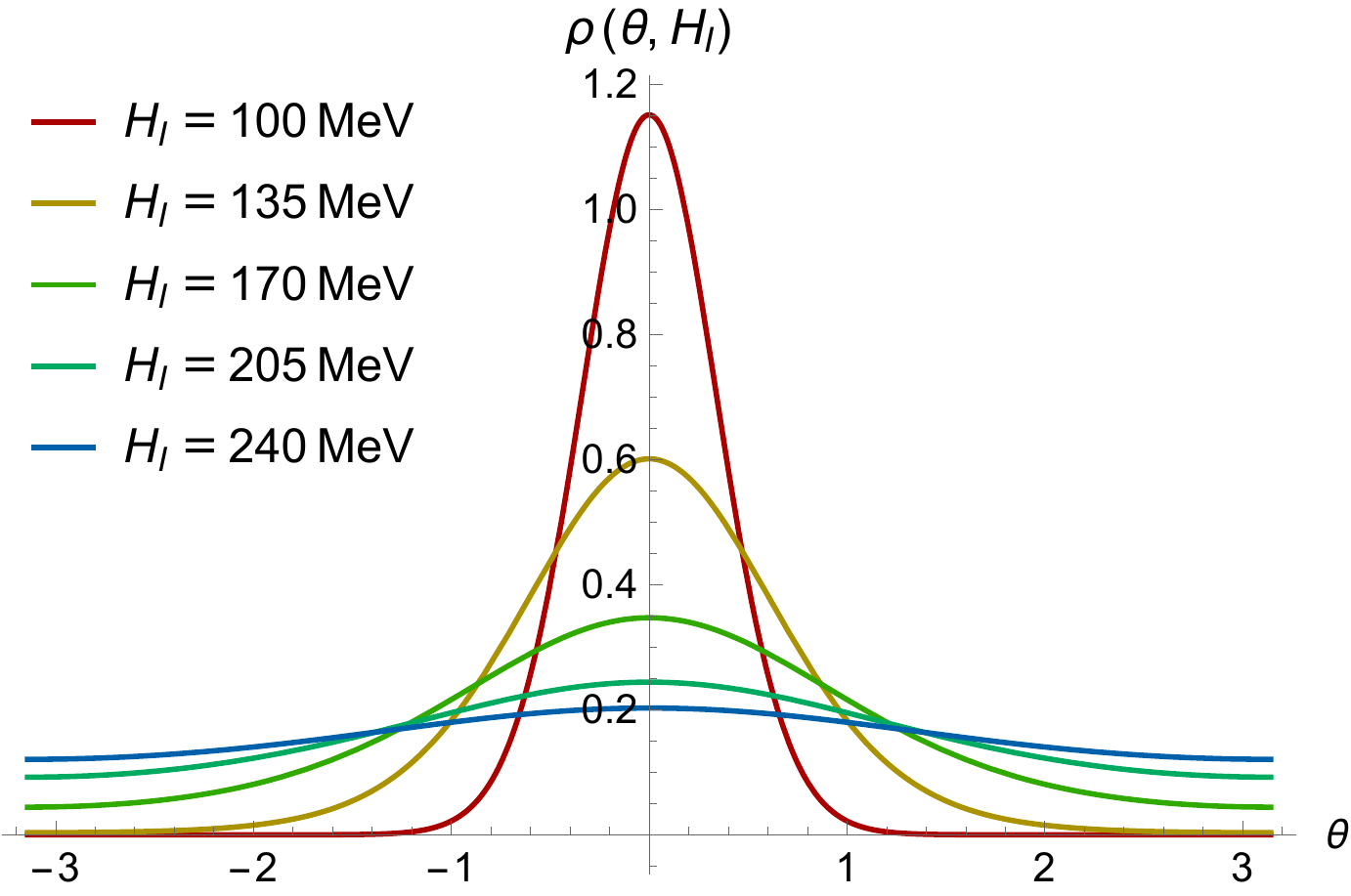}
	  \includegraphics[width=.45\textwidth]{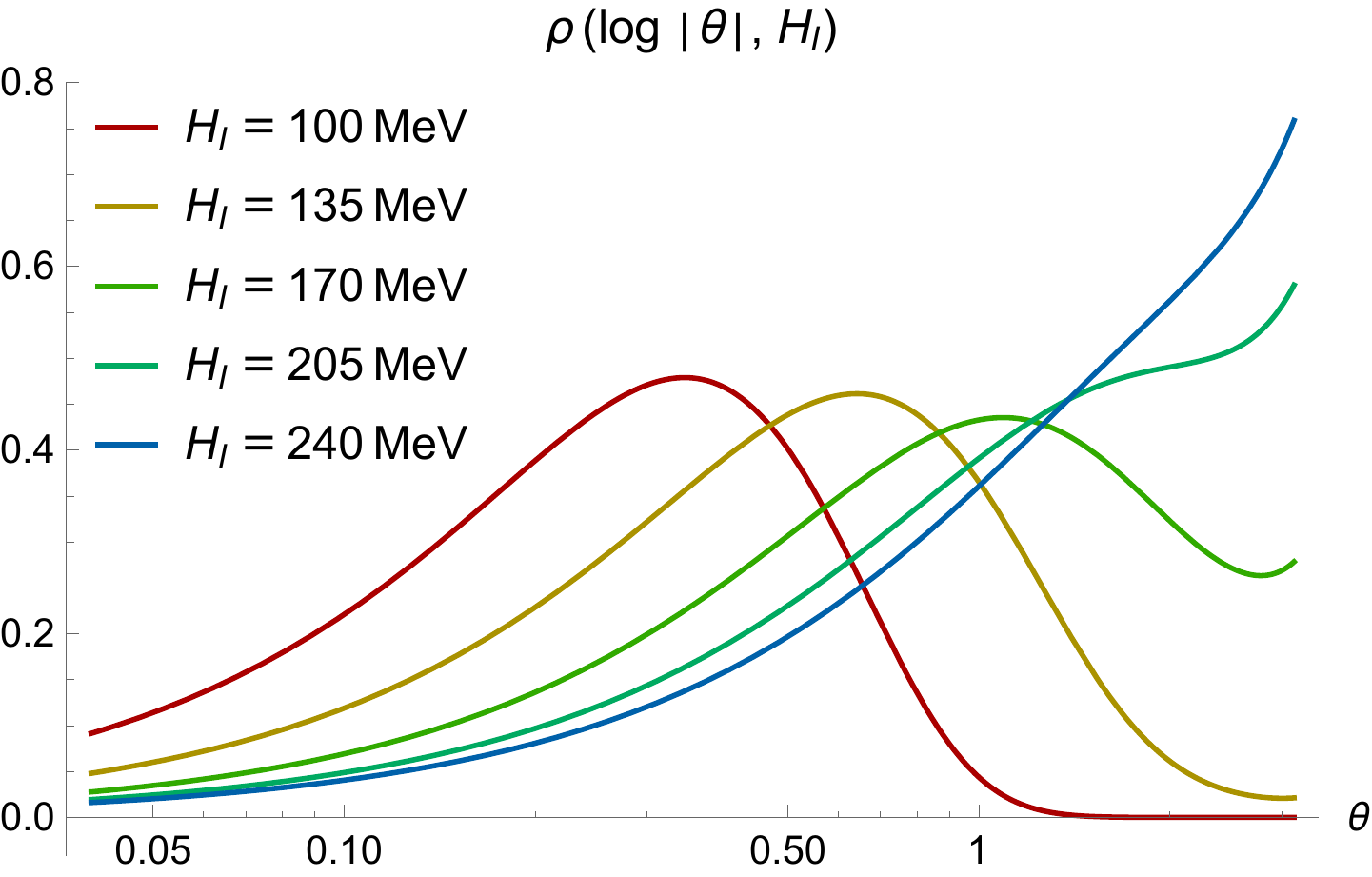}}
	\caption{Equilibrium distribution for the QCD axion, in terms of $\theta$ (left) and $\log|\theta|$ (right), at various $H_I$}
	\label{pdf}
\end{figure}

The exact equilibrium distribution for the axion is, up to normalization,
\eq{
\rho(\theta,H_I) &\propto \exp\left(-\frac{8\pi^2\chi(H_I/2\pi)(1-\cos\theta)}{3H_I^4}\right)
}
for $H_I\lesssim 2\pi T_c$, about $800\MeV$, the susceptibility is constant:
\eq{
\rho(\theta,H_I) &\propto \exp\left(-\frac{8\pi^2\chi(0)(1-\cos\theta)}{3H_I^4}\right)\\
&\approx \exp\left(-\frac{1-\cos\theta}{(H_I/170\MeV)^4}\right)
}
This distribution is known as the von Mises distribution. It is plotted in \Fig{pdf} for $H_I$ ranging from 100 to $240 \MeV$.

For $H_I\ll (8\pi^2\chi(0)/3)^{1/4}$, about $170\MeV$, this distribution is approximately a narrow Gaussian with width $(H_I/170\MeV)^2$:
\eq{
\rho(\theta,H_I) &\propto \exp\left(-\frac{\theta^2}{2(H_I/170\MeV)^4}\right)
}
The distribution $\rho(\log|\theta|)$ is also plotted. This distribution is peaked where $\theta$ is comparable in magnitude to the width of the original distribution. Note that, in the Gaussian (low $H_I$) regime, this distribution falls off slowly (power law in $\theta$) at small $\log|\theta|$, and quickly (exponentially in $\theta$) at high $\log|\theta|$. The left tail corresponds to the center of the Gaussian, where the distribution is smooth, with the power law coming from the change of measure on $\theta$; the right tail inherits its exponential suppression from the tails of the Gaussian.

This approximation breaks down as the width become $\mathcal O(1)$ and ``sees'' the anharmonic shape of the potential. For $H_I\gg 170\MeV$, the probability distribution becomes flat, with a small sinusoidal variation:
\eq{
\rho(\theta,H_I) &\propto 1-\frac{1-\cos\theta}{(H_I/170\MeV)^4}
}
The relative variation in $\rho$, across its range, is less than .001 for $H_I>2\pi T_c$, so we do not need to explicitly model the temperature-dependence of $\chi$ in this part of our analysis.

In general, we will refer to $H_I\lesssim 170\MeV$ as ``low $H_I$'' and $H_I\gtrsim 170\MeV$ as ``high $H_I$'' for the remainder of the paper.

As discussed in \Sec{rollout}, the axion density is determined by the value of $H$ at $\epsilon\sim m^2/H^2\implies \dot H\sim m^2$. This can be higher than the value of $H$ in the last 60 e-folds (the value observable in our universe).%, but for most low-scale inflation models there is negligible difference.\\

Because the axion density is a function of both the PQ scale $f_a$ and the initial (post-inflation) value of $\theta$, requiring axions to make up all of dark matter ($\Omega_a=\Omega_c$) imposes a relationship between $f_a$ and $\theta$. We will use $\theta_{DM}(f_a)$ to refer to the initial $\theta$ needed for a given $f_a$, subject to this constraint. For more details on the functional form of $\theta_{DM}(f_a)$, see \App{fth}.

\subsubsection{The Stochastic Axion Window}
For any value of $H_I$ and $f_a$, the value of $\theta$ in a given patch will be random, following the distribution $\rho$. In order to make our model natural, $\theta_{DM}(f_a)$ should not be an unusually high or unusually low value for this distribution. We can quantify this in terms of the probability $p$ that $|\theta|\le\theta_{DM}$, and the corresponding probability $q=1-p$ that $|\theta|>\theta_{DM}$. There are two ways for the observed $\theta$ to be ``unnatural'': it can be unnaturally large (very small $p$) or unnaturally small (very small $q$). Note that this is a very concrete notion of naturalness: we have a finite set of Hubble patches after inflation, with a known distribution of misalignment angles. In particular, $p$ is a fraction of Hubble patches, rather than an abstract or epistemic notion of probability.

Technically, $p(\theta)$ is the CDF of the distribution $\rho(|\theta|)$. In order to plot contours of $f_a$ and $H_I$ for fixed $p$, we need the inverse CDF $\theta(p)$. This is slow to compute numerically, particularly for extremely small values of $q$, but the Gaussian distribution for a noninteracting particle of the same mass is a good approximation in both limits. (For large $H_I$, a very wide Gaussian truncated at $|\theta|=\pi$ is approximately uniform, as desired). For values of $H_I$ around $144 \MeV$ we can interpolate numerically as long as $q$ is not too small.

\begin{figure}[h!]
	\centering
	  \makebox[\textwidth][c]{\includegraphics[width=\textwidth]{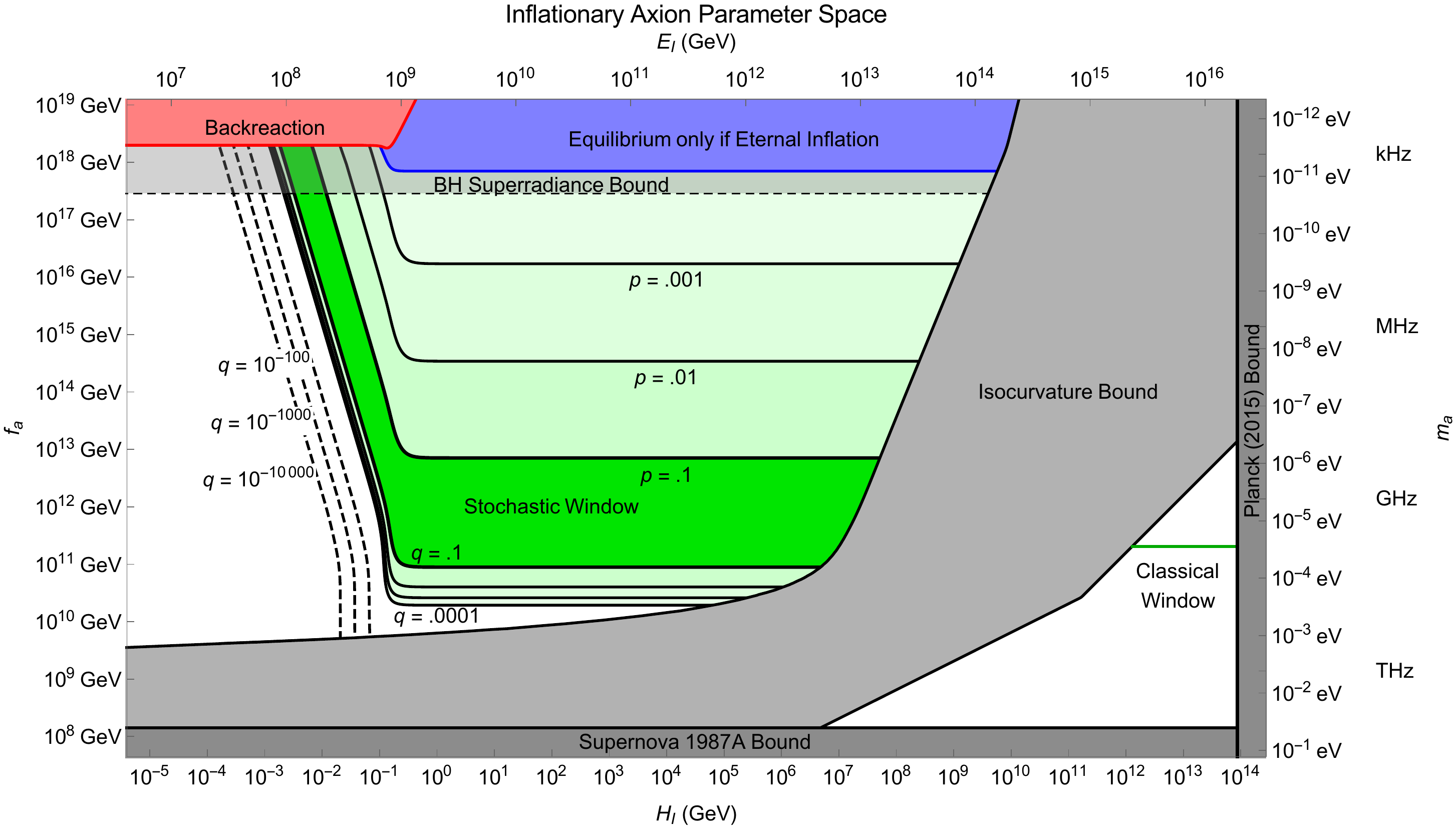}}
	\caption{Parameter space for the QCD axion dark matter, assuming a long enough period of inflation that the axion reaches equilibrium as described in the text. Axes are axion decay constant $f_a$ (left) and mass $m_a$ (right, inverted), Hubble scale of inflation $H_I$ (bottom), and inflationary energy scale $E_I=(3H_I^2M_P^2)^{1/4}$ (top).
\\
In the large green region, the observed dark matter density is a typical density to get from our axion equilibrium distribution ($p>.1$ and $q>.1$). Smaller values of $p$ and $q$ are shown as solid and dashed contours around this region. At near-Planckian $f_a$, the axion's behavior changes: in the pink region, backreaction effects become significant and force $\theta\to\pi$; in the blue region, the distribution does not reach equilibrium and depends on initial conditions, except in eternal inflation.
\\
At high $H_I$ and low $f_a$ is the classical window, where PQ symmetry breaks and produces axions after inflation. The thin green line shows the standard value of $f_a$ where this production matches the observed dark matter density.
\\
Observational constraints are shown in gray: isocurvature from the CMB spectrum, a lower bound on $f_a$ from supernova 1987A, black-hole superradiance, and an upper bound on $H_I$ from the Planck 2015 constraint on $r$.
}
	\label{bigplot}
\end{figure}

The axion parameter space is shown in \Fig{bigplot}. For high $f_a$, the value of Hubble during inflation depends on $f_a$ and is confined to about an order of magnitude (in order for $p$ to fall between 0.1 and 0.9). For higher $H_I$, it is still possible to obtain the correct dark matter density via anthropics, with fine-tuning ($p$) of no more than $10^{-4}$; for lower values of $H_I$, however, the value of $q$ drops off exponentially due to the tails of the Gaussian, and implausible amounts of fine-tuning are required to produce enough dark matter. Therefore, long periods of extremely-low-scale inflation are incompatible with the anthropic axion. The solid contours are for $p$ or $q$ equal to $10^{-1},10^{-2},10^{-3},10^{-4}$. The dashed contours are for $q=10^{-100},10^{-1,000},10^{-10,000}$, and use the Gaussian approximation without numerical interpolation. 

In short, we have carved up the former ``anthropic axion'' window into three regimes: for high $f$ and high $H_I$ we have a concrete model of an anthropic axion with a uniform prior probability, for very low $H_I$ we have a low axion abundance with $\theta$ extremely small, and in the middle there is a band where axion dark matter is naturally obtained, which we call the stochastic window. In this middle region (the stochastic window) $f_a$ ranges from $\approx 10^{11}\GeV$ to the Planck scale. Within this window, two observables become correlated: the Hubble scale of inflation and the mass of a dark-matter QCD axion. Lighter axions (higher $f_a$) require a lower Hubble scale.

The red region is ruled out due to a backreaction effect, which concentrates $\rho$ around the {\it maxima} of $V$, overproducing dark matter. This comes into play when the relaxation time for the axion is too long. This rules out a decay constant $f_a>\sqrt{2/3}M_P\approx 2.0\times 10^{18}\GeV$, or a mass $m_a\lesssim 2.9\times 10^{-12}\eV$, see \Sec{backrx} and \App{fokk}. The blue region is consistent with our model, but the relaxation time is longer than inflation can last without becoming eternal; in this window, either inflation is eternal or it is so short that the misalignment angle is set by initial conditions. Also shown are observational constraints due to isocurvature~\cite{Kobayashi:2013nva,diCortona:2015ldu} and supernova 1987A~\cite{Raffelt:2006cw,Patrignani:2016xqp,Chang:2018rso}. The gray region shown at high $f_a$ is disfavored due to black-hole superradiance~\cite{Arvanitaki:2014wva}. There is an upper bound on $H_I$ (during the last 60 e-folds) from Planck and BICEP2/Keck constraints on primordial $B$-modes~\cite{Ade:2015tva,Ade:2015lrj}.

For other analyses of $(H_I,f_a)$ parameter space, see e.g.~\cite{Wantz:2009it,Kobayashi:2013nva,Visinelli:2009zm,Visinelli:2014twa,diCortona:2015ldu,Steffen:2008qp,Beltran:2006sq,Diez-Tejedor:2017ivd,Hamann:2009yf,Kawasaki:2018qwp,Visinelli:2017imh,Ballesteros:2016xej}.  Our results agree with these, up to the fact that we have added the allowed dark matter region to the left of the isocurvature bound of course.

\begin{figure}[h!]
	\centering
	  \makebox[\textwidth][c]{\includegraphics[width=.495\textwidth]{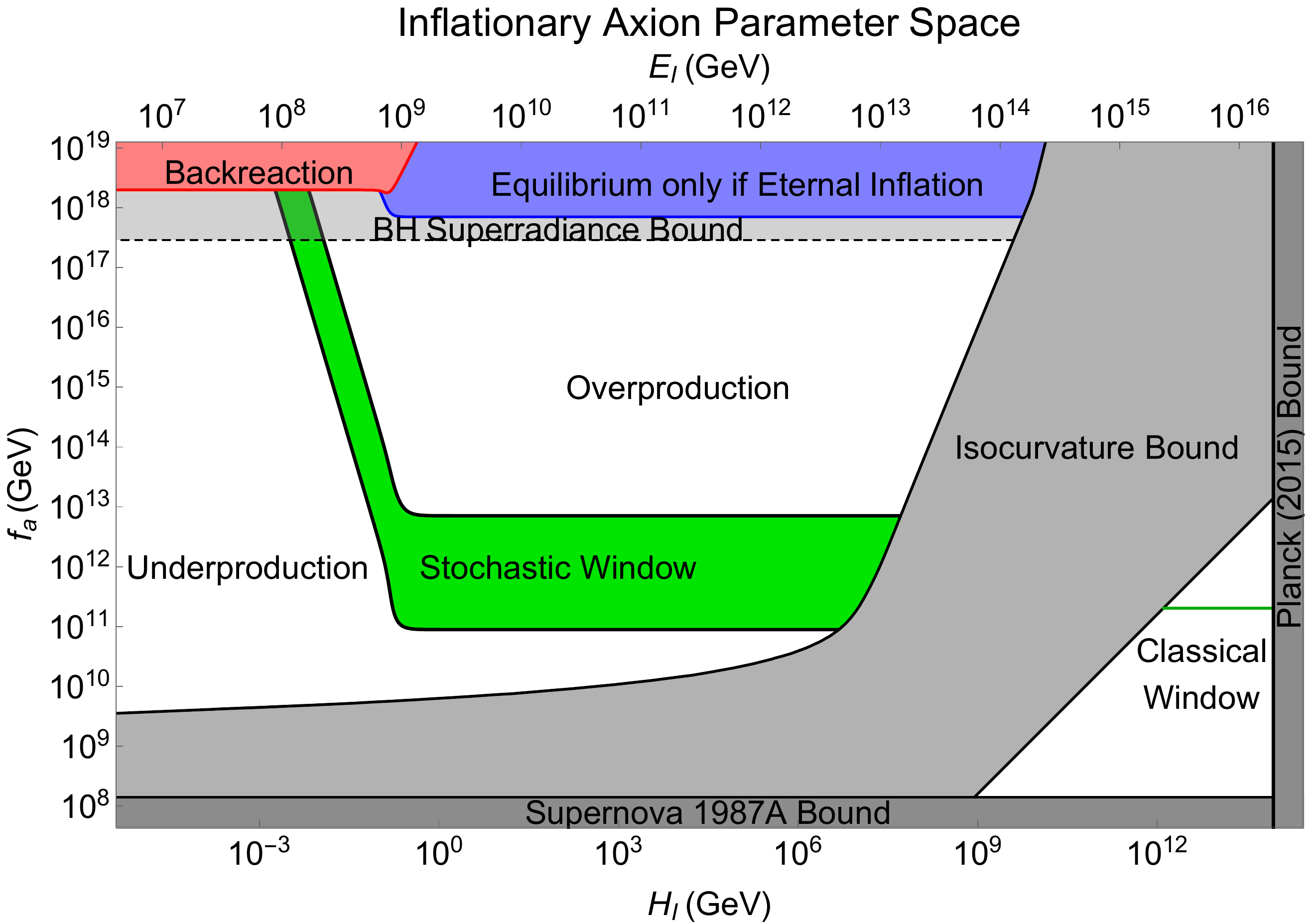}
	  \includegraphics[width=.5\textwidth]{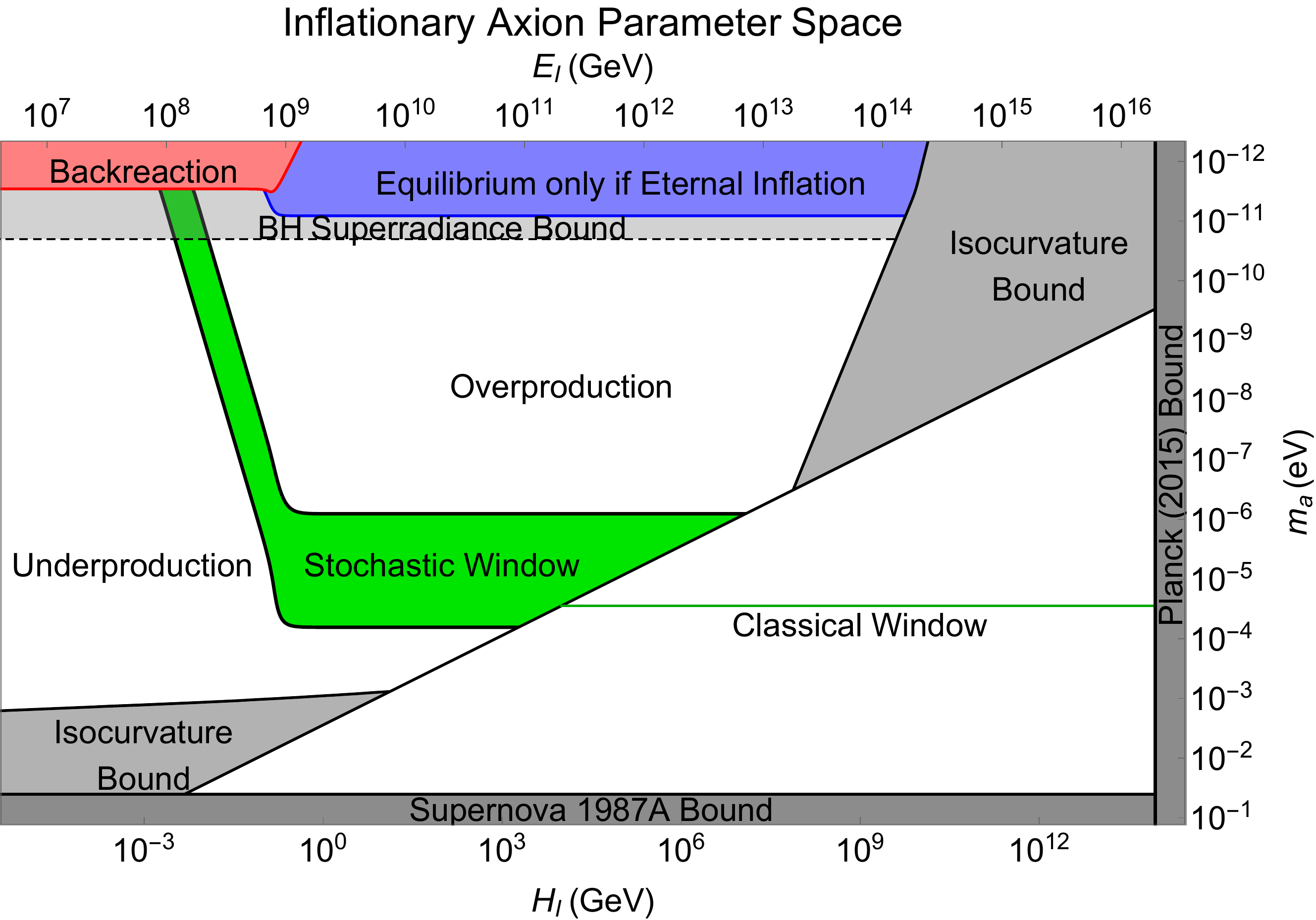}}
	\caption{The same parameter space as \Fig{bigplot} except with maximally inefficient (left) or maximally efficient (right) reheating.  The main difference is that the isocurvature bound moves and so does the boundary between the classical and stochastic windows. If reheating is inefficient, then the axion is produced after inflation if $T_{dS}=H_I/2\pi>f_a$. If it is maximally efficient, then the weaker bound $E_I>f_a$ holds instead. In general, the bound is given by $\max(T_{dS},T_{rh})>f_a$ where $T_{rh}=\epsilon_{eff}E_I$ is the reheating temperature.}
	\label{smallplot}
\end{figure}

\subsubsection{Classical Window}
The fate of the axion depends on whether the universe reheats to a temperature above $f_a$. The reheating temperature depends on the efficiency of reheating, $\epsilon_{eff}\leq1$:
\begin{align}
T_{rh}=\epsilon_{eff}E_I
\end{align}
where $E_I=(3H_I^2M_P^2)^{1/4}$ is the energy scale of inflation.

If $T_{rh}>f_a$, PQ symmetry is restored, the misalignment angle is destroyed, and axions are produced through a different mechanism when the universe cools and PQ breaking occurs. Even if reheating does not reach this temperature, the de Sitter temperature during inflation could, if $H_I/2\pi>f_a$. In this case, PQ symmetry is maintained throughout inflation, so no axions and no isocurvature fluctuations are produced to begin with. These two lines form the right edge of the ``Isocurvature Bound'' region in \Fig{bigplot}, with $\epsilon_{eff}=10^{-4.5}$ for the sake of illustration. The same parameter space is shown in \Fig{smallplot} with $\epsilon_{eff}\ll 1$ (very inefficient reheating) and $\epsilon_{eff}=1$ (maximally efficient reheating).

In the region to the right of this boundary, PQ symmetry is preserved or restored until after reheating. Once the universe cools below $f_a$, the process of symmetry breaking randomizes the axion and gives it an effective misalignment angle $\theta_C$ everywhere. This implies a definite value $f_a=f_C$ for which there is the correct abundance of dark matter ($\theta_C=\theta_{DM}(f_C)$), which we refer to as the classical axion window and show as the thin green line on the right side in \Fig{bigplot}. There is significant systematic uncertainty about this value, due in part to the difficulty of accounting for axion string decay contributions, which increase $\theta_C$.

%We assume an efficiency of $\epsilon_{eff}=10^{-4.5}$ for reheating, for the sake of illustration, and show this boundary as $\epsilon_{eff}^43M_P^2H_I^2>f_a^4$ on the plot. (To see how this changes at various $\epsilon_{eff}$, see \cite{Steffen:2008qp}, Fig. 2.)  On the right side of the line, isocurvature constraints are no longer produced.

Suppose $f_a$ is slightly higher than $f_C$. Naively, the correct dark matter abundance could still be obtained after PQ symmetry breaking if the axion accidentally ended up near the (eventual) minimum of its potential by random chance. However, this feat would need to be replicated independently in many causally-disconnected Hubble patches; a single Hubble patch today contains over $10^{50}$ regions that were Hubble patches at PQ breaking. Therefore, the probability of an effective misalignment angle less than $\theta_C/10$, in each of these patches, is $10^{-10^{50}}$. This is an absurdly small number even compared to the ``dead'' axion at very low $H_I$. The same conclusion applies in the other direction, with $f_a<f_C$; the classical axion window is extremely narrow.

\subsection{Fuzzy Dark Matter}

This same stochastic mechanism for producing dark matter abundance from quantum fluctuations of a field during inflation can apply more broadly than just to the QCD axion.  In this section we consider a general scalar field, and in particular motivated by fuzzy dark matter (see e.g.~\cite{Hui:2016ltb}).

\subsubsection{Free Scalar Model}
For the case of a free scalar field, our mechanism works to produce the observed dark matter density over a wide range of masses. The required Hubble scale is roughly
\eq{H_I&\approx 500 \eV \left(\frac{m}{10^{-22} \eV}\right)^{3/8}\\
&=500 \GeV \left(\frac{m}{100 \eV}\right)^{3/8}}

There is a lower bound on mass from the backreaction and eternal-inflation bounds (see \Sec{inflation}), which are {\it both} parametrically $H_I^2\lesssim mM_P$ in this case. This rules out $m\lesssim10^{-26}\eV$, but also cuts into the stochastic window for any $m\lesssim 10^{-22}\eV$. In this case, the backreaction behaves differently than for axions; rather than sitting on a hilltop, the field runs away to extremely large field values, stopping only when it backreacts significantly on spacetime and begins to act as an inflaton.

The upper bound is set by isocurvature (without anharmonic corrections), which rules out masses greater than around $100\MeV$. At this upper bound, $H_I\approx50\TeV$.

This leaves essentially the entire range where weakly-coupled scalar dark matter is of interest. In particular, fuzzy dark matter $(m\approx10^{-22}\eV)$ is allowed. The observed abundance of dark matter can be reached naturally for a quadratic fuzzy dark matter (FDM) potential with $H_I$ of $200-800\eV$. This is the stochastic window for FDM (but $H_I>600\eV$ requires eternal inflation to reach equilibrium).

\subsubsection{Axionic Model}
For fuzzy dark matter, one common assumption is a non-QCD axion with decay constant $f_a\approx 10^{17}\GeV$ and a uniformly random initial $\theta$~\cite{Hui:2016ltb}.

For an axion-like potential, we can easily accommodate higher $f_a$ by adjusting $H_I$ downwards slightly to produce a small $\theta$; on the other hand, for $f_a\approx 10^{17}\GeV$, any $H_I\gtrsim1\keV$ will give a uniformly random initial $\theta$. The usual backreaction bound for axions, $f_a\lesssim M_P$, applies in this case; the backreaction bound for a free scalar assumes that the maximum height of the potential is comparable to the inflaton. Therefore, large values of $H_I$ are permitted, up to the isocurvature limit.

For even lighter axion-like particles, the spatial variation of the ALP field (caused by inflationary fluctuations in the last 60 e-folds) can rotate CMB $E$ modes into $B$ modes. This places a bound on the ALP-photon coupling at these very low masses~\cite{Pospelov:2008gg,Array:2017rlf}.

\section{Inflationary Sector}
\label{inflation}
In the analysis above, we have made several simplifying assumptions about the background spacetime. In essence, we have been working with de Sitter -- an eternally-expanding non-dynamical background geometry with constant $H$ -- rather than an inflationary FLRW universe, where $H(t)$ changes, inflation may last for only a finite duration, and the spectator field may backreact on the spacetime geometry. Taking this all into account produces several modifications of the picture above.

First, the $H_I$ we discuss above is not necessarily the value of $H$ observed from the last 60 e-folds, and can be significantly higher. A freeze-out-like mechanism, based on the slow-roll parameter rather than temperature, sets the applicable $H_I$.

Second, inflation may not last long enough for the dark matter field to reach its equilibrium distribution. We calculate the number of e-folds required for the QCD axion, and find a regime where the relaxation time is longer than can be achieved without eternal inflation. If inflation does not last long enough for significant relaxation, the dark matter density is set by initial conditions.

Finally, the dark matter field may backreact. In particular, patches with higher $V(\phi)$ will expand at a slightly faster $H$, which allows them to outnumber the lower-$H$ patches if the relaxation time is long enough. We find that this occurs for low $H_I$ and nearly-Planckian $f_a$ for the QCD axion; in this regime, the backreaction effect pushes the axion towards $\theta=\pi$, where the faster exponential growth of space outpaces the relaxation process.

\subsection{Variation in $H_I$}
\label{rollout}
In this section, we drop the assumption that $H_I$ is constant over time.

The equilibrium distribution for a massive scalar gives an expectation value for the field's energy density, $\braket{V(\phi)}$, of order $H_I^4$. The relaxation time is of order $H_I/m^2$, or $H_I^2/m^2$ e-folds. In an inflating universe, as opposed to de Sitter, $H$ is a function of time. In order for the field to approach equilibrium faster than the equilibrium density is changing, we need
\begin{align}
\frac{d\log H}{dt}&\lesssim \frac {m^2}H\\
\dot H&\lesssim m^2\\
\epsilon&\lesssim \frac{m^2}{H^2}
\end{align}
where $\epsilon=\dot H/H^2$ is the first slow-roll parameter. Therefore, the field stays in equilibrium as long as the inflaton is rolling slowly enough, $\epsilon\ll \frac{m^2}{H^2}$, and ``freezes out'' at $\epsilon\approx\frac{m^2}{H^2}$. The energy density of the field after reheating will be of order $H_*^4$, where $H_*$ is the Hubble scale at the last time that $\dot H\lesssim m^2$. (This may occur multiple times, because $\dot H$ doesn't have to change monotonically.) Any field with $m^2$ below the minimum value of $\dot H$ will be ``frozen'' the entire time and remain at its initial field value. If inflation is never chaotic, then $\dot H>H^4/M_P^2$ throughout, so any field with $m<H_{\text{rh}}^2/M_P$ will fall into this category. %On the other hand, any field with $m>H_{\text{rh}}$, the Hubble scale at reheating, will remain in equilibrium throughout and will have density of order $H_{\text{rh}}^4$. %but it'll roll! so never mind.
%there's a cute diagram I could draw and put here explaining this all a bit better graphically

Consider the case of multiple scalar fields. Because $H$ decreases monotonically, the lightest weakly-coupled scalar will have the highest density after reheating. In many low-scale inflation models (such as hilltop models) $H$ is nearly constant, so the densities at reheating will be of the same order of magnitude (the exact densities are random). Light fields begin to dilute later, giving an extra factor of $m^{-3/2}$ to the density. (This assumes that all of the scalars have masses above $H_{\text{eq}}\approx10^{-28}\eV$, the Hubble at matter-radiation equality, and that their potentials do not change after oscillation begins). Therefore, the lightest scalar will almost always be the dominant component; if the second-lightest is only a few times heavier, however, it is not too unlikely for it to have a higher density through random chance. 

If there are a sufficient number of scalars over a range of masses whose abundances can be measured, then this $m^{-3/2}$ dependence could provide an observable signature that distinguishes the stochastic scenario from anthropic selection, which generically causes all scalars to have comparable densities (except for scalars that do not have overclosure problems in the first place, e.g. axions with low $f_a$).

\subsection{Length of Inflation}
In this section, we discuss how long inflation needs to last for $\rho$ to achieve equilibrium.

For $H_I\lesssim 170\MeV$, as shown in \Sec{QCD}, we can treat the axion as a free scalar. In this case, the equilibrium distribution is reached on a timescale of $3H_I/m^2$. Therefore, for low $H_I$, the production mechanism and parameter-space analysis above are relevant when the number of e-folds is $N\gg 3H_I^2/m^2$, where $N_r=3H_I^2/m^2$ is the relaxation time in e-folds. For a QCD axion, with $H_I$ in the $\MeV-\GeV$ range, this is anywhere from about $10^{20}$ to $10^{40}$ e-folds. This corresponds to a relaxation time ranging from one week (at $f_a=10^{13}\GeV$) to several million years (at $f_a\approx M_P$). While much higher than the 50-60 e-folds required by the horizon problem, this is  achievable given a flat enough potential.  Note that models such as the relaxion to solve the hierarchy problem require a similarly low Hubble scale and even longer periods of inflation (see e.g.~\cite{Graham:2015cka}), which can provide some motivation for considering such inflationary sectors.

For $H_I\gtrsim 170\MeV$, we should treat the potential as flat, and need to take the axion's compact range into account. As noted in \Sec{compact} and \App{fokk}, the relaxation time in e-folds is $N_r=8\pi^2f_a^2/H_I^2$ 
In this regime, the stochastic window is restricted to $f_a\lesssim10^{14}\GeV$, so $N_r$ is at most $10^{30}$ or so.

These relaxation times place a lower bound on the number of e-folds $N\gtrsim N_r$ necessary for our model to be independent of initial conditions. This can be converted to a slow-roll parameter $\epsilon$. This parameter can be expressed as $-d\log H/dN$, so the dark matter field is in equilibrium as long as, at some point during inflation, $\epsilon$ remains $\lesssim1/N_r$ over at least an order of magnitude in $H$.

There is also an upper bound on $N$ for any non-eternal inflationary model, and an upper bound on $N_r$ which avoids a backreaction effect. These are discussed in the following two sections.

\subsection{Eternal Inflation}
One way to obtain very long periods of inflation is via eternal inflation. This gives rise to an infinite volume of the universe, with accompanying measure problems. If we avoid eternal inflation, then an initial Hubble patch evolves into a finite population of Hubble patches at reheating, and the distribution $\rho(\theta)$ of misalignment angles can be interpreted more easily.

There is actually a sharp upper bound on the number of e-folds of inflation that can be obtained while still remaining in the non-eternal regime, given by\cite{Dubovsky:2011uy}
\eq{
N < \frac{2\pi^2M_P^2}{3H_I^2}
}
We derive this from a related bound in \App{clroll}.

Given this constraint, in the free scalar case, it is possible to maintain classical rolling for enough e-folds to reach equilibrium as long as $3H_I^2/m^2< 2\pi^2M_P^2/3H_I^2$, which is equivalent to
\eq{m &> \frac{3H_I^2}{\sqrt2 \pi M_P}}
For the QCD axion at low $H$, this implies
\eq{\frac{f_a}{M_P} < \frac{\sqrt2}3\frac{\Lambda_{QCD}^2}{H^2}
}
where $\Lambda_{QCD}=\chi(0)^{1/4}$. This constraint is satisfied in the $H_I\lesssim170\MeV$ branch of the stochastic window.

For higher $H_I$, our constraint is instead $8\pi^2f_a^2/H_I^2< 2\pi^2M_P^2/3H_I^2$, which is equivalent to
\eq{\frac{f_a}{M_P} &< \frac1{2\sqrt3}
}
This is always true in the $H_I\gtrsim170\MeV$ branch of the stochastic window. Note that this is of order $M_P$ like the backreaction bound, but for high $H_I$ instead of low $H_I$; together, they rule out transPlanckian or nearly-Planckian $f_a$ in the case of non-eternal inflation.

In the region above the stochastic window, these two constraints are not always obeyed. We can compute a more precise constraint, valid for all $H_I$ including $H_I\approx170\MeV$, by finding the fastest nonzero decay rate of a quasinormal mode of the Fokker-Planck equation (see \App{fokk}); we do this numerically, and confirm that it matches our analytical results in both regimes. The result is the region labeled ``Eternal inflation'' in \Fig{bigplot}; within this region, inflation must either violate the classical-rolling constraint (leading to a finite probability of an infinite reheating volume, i.e., eternal inflation) or have a duration of less than one relaxation time, in which case our results do not apply and the misalignment angle is simply determined by initial conditions prior to inflation. In either case, our concrete notion of fine-tuning -- based on a finite set of Hubble patches for which the distribution of field values is calculable -- is inapplicable.

\subsection{Backreaction}
\label{backrx}
In this section, we drop the assumption that the field does not backreact significantly on spacetime and find in what region backreaction is significant and cannot be neglected.

The height of the axion potential, $\chi\approx\Lambda_{QCD}^4$, is much smaller than the inflaton's energy density $3H_I^2M_P^2$ for our entire parameter space. However, this does not mean that backreaction effects can be ignored. Although the change in Hubble scale $\Delta H$ between $\theta=0$ and $\theta=\pi$ is tiny, it can add up significantly over a long relaxation time $t_r$: the number of patches with $\theta\approx\pi$ will be enhanced by up to $\exp(3\Delta H t_r)$. Over several relaxation times, each patch experiences the same distribution of values for $\theta$ and $H$, so this effect does not compound for much longer than $t_r$.

We can get a quick parametric estimate of when $3\Delta H t_r \gtrsim1$ fairly easily. As shown in \App{fokk}, $3\Delta H \approx V_{\text{max}}(\phi)/(2M_P^2H_I)\approx\chi/(2M_P^2H_I)$. Therefore, our bound is roughly
\eq{N_r \lesssim \frac{2M_P^2H_I^2}{\chi}}
The total number of e-folds of inflation can be larger than this; it is a bound only on the relaxation time.

For low $H_I$, we have $t_r\approx 3H_I/m_a^2$, so we should be concerned when $3\Delta Ht_r\approx3\chi/(2M_P^2m_a^2)\gtrsim1$. We have $\chi\approx m_a^2f_a^2$ for low $H_I$, so backreaction effects impose a bound on $f_a$ which is roughly
\eq{f_a\lesssim \frac{\sqrt2}{\sqrt3}M_P \label{eqn:backreaction1}}

For $H_I\gtrsim 170\MeV$, we have $t_r\approx 8\pi^2f_a^2/H_I^3$, so that $3\Delta Ht_r\approx \chi4\pi^2f_a^2/(M_P^2H_I^4)$. For $H_I\lesssim800\MeV$, $\chi\approx \chi(0)\approx m_a^2f_a^2$, so backreaction effects kick in around $(\chi(0)4\pi^2f_a^2)/(M_P^2H_I^4)\approx1$, which imposes the bound
\eq{\left(\frac{3f_a^2}{2M_P^2}\right)&\lesssim\left(\frac{H_I}{170\MeV}\right)^4 \label{eqn:backreaction2}}
As $H_I$ rises above $170\MeV$, this bound rapidly passes the Planck scale. By the time we reach $H_I\approx800\MeV$, where $\chi$ begins to decrease, the backreaction effects are negligible for any $f_a\lesssim 20M_P$; the decrease in $\chi$ makes them weaker still, so we can again ignore the temperature-dependent axion mass.

For a more detailed analysis of backreaction effects, which confirms these rough estimates and gives them more precise meaning, see \App{fokk}. In \Fig{bigplot}, the region ruled out by backreaction is plotted by setting $N=1/t_r$ and calculating $t_r$ numerically via the eigenvalue method described in \App{fokk}.  This  roughly agrees with the two approximate limits in their respective regions from equations \eqref{eqn:backreaction1} and \eqref{eqn:backreaction2}.  In this region, for a long enough period of inflation, axion dark matter is significantly overproduced.

\section{Conclusions}

We have found that, if inflation happens at a low scale, the minimal QCD axion is naturally driven towards an equilibrium distribution during inflation.  This equilibrium is between the classical force driving the axion to the minimum of its potential and the stochastic quantum fluctuations of the axion field.  With a sufficiently long period of such low-scale inflation, the axion will reach its equilibrium.  This equilibrium is independent of the precise length of inflation and depends only on the axion mass $m_a$ and the Hubble scale of inflation $H_I$.  Then we can predict the approximate size (officially the probability distribution) of the `initial' axion misalignment angle $\theta$ after inflation and thus the final axion abundance today.  This equilibrium prediction is shown in \Fig{bigplot}.  The region in which the axion achieves roughly the measured dark matter abundance is shown as the broad green band.  In the classical axion window (the thin green line on the right of the figure), the axion abundance is determined precisely by the axion mass.  In our new region the axion mass and $H_I$ determine only the rough size of the axion abundance but it can still vary by $\mathcal{O}(1)$ (part of why the region is broad).  Thus we can see that the minimal QCD axion model can naturally produce the correct dark matter abundance for a wide range of axion masses from roughly $10^{-4} \eV$, a little above the classical axion window, all the way down to the lowest mass around $10^{-12} \eV$.  This corresponds to axion decay constants from $f_a \sim 10^{11}\GeV$ all the way up to almost the Planck scale.

This axion equilibrium will arise from a sufficiently long period of normal inflation, or could also arise if there was a period of eternal inflation in our past (e.g.~\cite{Linde:1986fd}).  Eternal inflation (followed perhaps by some period of normal inflation) would certainly be long enough to produce an equilibrium distribution for the axion, however it also famously comes with measure problems and so it might not be possible to use our predicted probability distribution for the axion.

Interestingly, in our mechanism the axion abundance arises from the stochastic quantum fluctuations of the axion field during inflation, but nevertheless there are no observable isocurvature fluctuations induced.  The reason is that over a long period of inflation the axion field spreads out significantly in its potential well, but each individual quantum `jump' of the field is quite small.  Thus, in the last 60 or less e-folds (those relevant for the observable universe today) the axion spreads out only a very little distance in field space in its potential and so the isocurvature perturbations of wavelengths smaller than today's horizon size are quite small.  This same mechanism can also produce a nonzero abundance of other scalar fields (e.g.~for fuzzy dark matter) from quantum fluctuations during inflation without producing dangerous isocurvature perturbations.

Thus we see that just the minimal QCD axion model can naturally reproduce the observed dark matter abundance down even to the lowest masses ($f_a$ almost up to the Planck scale).  This motivates searching experimentally for QCD axions broadly over the entire mass range.

\section*{Acknowledgements}
We acknowledge useful conversations with Savas Dimopoulos, Michael Dine, David E.~Kaplan, Andrei Linde, Surjeet Rajendran, Leonardo Senatore, and Eva Silverstein.  This work was supported by NSF grant PHY-1720397, DOE Early Career Award DE-SC0012012, Heising-Simons Foundation grant 2015-037, and the Mellam Graduate Fellowship.
Note added: Shortly after this paper appeared, another overlapping paper appeared which generally agrees with our results \cite{Guth:2018hsa}.

\appendix

\section{Misalignment Angle and PQ Scale $f_a$}
\label{fth}

Depending on $f_a$, there are three regimes for the behavior of the misalignment mechanism for a QCD axion. In all cases, the axion is frozen by Hubble friction until $3H\lesssim m_a$, after which it oscillates and behaves as cold dark matter. The misalignment angle needed for a correct abundance of axion dark matter, $\theta_{DM}(f_a)$, can be calculated by extrapolating the axion density backwards from the present to the time of oscillation. $\theta_{DM}$ is a roughly power-law function of $f_a$, ranging from about $10^{-4}$ at $f_a\sim M_P$ to $\mathcal O(1)$ for $f_a\sim10^{12}\GeV$.

At high $f_a$ (low $m_a$), the axion is frozen until after the QCD phase transition. In this case, the density now is just the density at oscillation, diluted by a factor of $a^3$: 
\eq{\Omega_a \sim m_a^2\phi^2a_{\text{osc}}^3}
In a radiation-dominated universe, $a\propto H^{-1/2}$, so the density is proportional to $\Lambda_{QCD}^2\theta^2H^{-3/2}$. Noting that $H_{\text{osc}}\sim m_{a} \sim f_a^{-1}$, the density goes as $\theta^2 f_a^{3/2}$. Holding this constant, we have that $\theta_{DM}\propto f_a^{-3/4}$.

At lower $f_a\lesssim 10^{17}\GeV$, when the axion oscillates before the QCD transition, the mass at the time of oscillation $m_{a,\text{osc}}$ is a power-law function of the temperature $T_{\text{osc}}$ of the quark-gluon plasma. This introduces an additional factor of $f_a^n$ into $\theta_{DM}$, where the exponent $n$ can be derived analytically (e.g. dilute instanton gas approximation (DIGA)) or numerically via lattice methods (e.g.~\cite{Petreczky:2016vrs,Berkowitz:2015aua,Borsanyi:2016ksw}). For example, DIGA (with three light quarks, to leading order) gives $m\propto T^{-4}$ and $\theta_{DM}\propto f_a^{-7/12}$~\cite{Petreczky:2016vrs}. However, we do not use DIGA. Instead, we combine results from the more accurate numerical studies, which produce various exponents $n$ in the vicinity of $-7/12$.
There is a wiggly bend in $\theta_{DM}$ at $f_a\approx 10^{17}\GeV$, corresponding to a dark-matter axion which begins to oscillate at the QCD transition.

At even lower $f_a\approx 10^{12}\GeV$, the axion must begin with a misalignment angle close to $\pi$. At this point, anharmonic corrections must be included; one paper\cite{Kobayashi:2013nva} gives the functional form in this regime as $\theta_{DM} = \pi-\exp(Af_a+B)$ for some constants $A,B$.

To get a more precise model for $\theta_{DM}$, we fit a nonlinear function to the $f_a$ vs. $\theta_{DM}$ curves from several papers which use various techniques to deal with these three regimes\cite{Wantz:2009it,Kobayashi:2013nva,Borsanyi:2016ksw}. There is a disagreement in the literature regarding the overall prefactor for $f_a$ in this relation, so we multiplied the values of $f_a$ from \cite{Wantz:2009it} by $1.133$ and from \cite{Kobayashi:2013nva} by $1.676$ to agree with \cite{Borsanyi:2016ksw} (the most recent work) in the region of overlap of the curves. (These values were obtained from a numerical fit.)

First, we transformed $\theta$ nonlinearly to a new variable
\eq{
X &= \left(\frac\theta\pi\right)\left(1-C\frac\theta\pi-\log\left(\frac{\pi-\theta}\pi\right)\right)
}
where the constant $C\approx 1.267$ was fit by hand. $X$ is linear in $\theta$ at $\theta\ll1$, and logarithmic in $\pi-\theta$ at $\theta\approx\pi$. We fit a smoothed piecewise power law for $f_a(X_{DM})$, with a power of $-4/3$ for high $f_a$, $-1$ for low $f_a$, and an arbitrary power (the fit produced $-1.699$, close to $-12/7$ as expected) in between.

This translates into an analytic approximation for $f_a(\theta_{DM})$ which is power-law at small angles, with a change in the exponent around $10^{18}\GeV$ (which fits better than $10^{17}$ due to the ``wiggliness'' of the transition) and linear in $\log(\pi-\theta)$ at low $f_a$, which accounts for anharmonic effects when the axion is on the ``hilltop'' at $\theta\approx\pi$\cite{Kobayashi:2013nva}.

The fitted function agrees well with the (shifted) curves from \cite{Wantz:2009it,Kobayashi:2013nva,Borsanyi:2016ksw}: within 5\% except where the source papers disagree by about 20\% near the $10^{17}\GeV$ transition.

\section{Fokker-Planck Formalism and Inflationary Backreaction}
\label{fokk}
This appendix follows the approach of \cite{Stopyra:2014}, specialized to the QCD axion potential and with the additional introduction of the backreaction effect.

Consider a slow-rolling scalar field $\phi$ in an expanding spacetime. As in the body of the paper, we consider the long-wavelength modes separately from the short-wavelength modes and focus on the former. % details: \phi below is Stopyra's phi-bar, the average over a Hubble volume around some point. If we fix that point to be comoving then we don't need to worry about the spatial part of f(x,t).
Starobinsky\cite{Starobinsky:1994bd} showed that these evolve classically, according to the usual slow-roll equation with an additional stochastic (random walk) term:
\begin{align}
\dot\phi=-\frac{V'(\phi)}{3H_I}+f(t)
\end{align}
where $f_a$ is Gaussian noise with correlation function
\begin{align}
\braket{f(t_1)f(t_2)}=\frac{H_I^3}{4\pi^2}\delta(t_1-t_2).
\end{align}
This is a Langevin equation, which describes the evolution of $\phi$ over time as a stochastic random variable. It is more convenient to work in terms of the probability density $\rho(\phi,t)$, which gives us a (deterministic) Fokker-Planck equation:
\begin{align}
\dot\rho(\phi,t) &= \frac1{3H_I}\partial_\phi(V'(\phi)\rho(\phi,t))+\frac{H_I^3}{8\pi^2}\partial^2_{\phi\phi}\rho(\phi,t)
\end{align}
Any initial probability distribution will asymptotically approach an equilibrium distribution of the form:
\begin{align}
\rho_f(\phi,t) \propto \exp\left(-\frac{8\pi^2V(\phi)}{3H_I^4}\right)
\end{align}

%overview
In the body of the paper, we derive parametric estimates of the relaxation time for the axion distribution $\rho$ and of the scale $f_a$ where backreaction effects become significant, in the small and large $H_I$ regimes. We can obtain more precise descriptions of these quantities by decomposing the time evolution of $\rho$ into quasinormal modes. The shapes and half-lives of these modes are the eigenfunctions and eigenvalues of a Schr\"odinger-like equation with a potential that is related to the axion potential.

To understand the backreaction effect, we will momentarily work in terms of
\eq{
P(\phi,t) &= e^{3Ht}\rho(\phi,t)
}
This distribution counts the total number of patches with average field value $\phi$, instead of the relative frequency. Whereas the integral $\int\rho\du\phi$ is always 1, the integral of $P$ grows as the universe expands. We can write a Fokker-Planck equation for $P$ by including a term for this growth:
\eq{
\dot P(\phi,t) &= \frac1{3H}\partial_\phi(V'(\phi)P(\phi,t))+\frac{H^3}{8\pi^2}\partial^2_{\phi\phi}P(\phi,t)
+3H P(\phi,t)
}
However, $H$ is not truly independent of $\phi$. Patches where the field is farther up its potential will have a slightly higher energy density $V_{\text{tot}}=V_I+V(\phi)$, where $V_I$ is the energy density due to the inflaton and $V$ is the potential for $\phi$. (Of course, this decomposition depends on our choice of zero for $V(\phi)$; we choose $V(0)=0$, so that $V(\phi)\ge0$.) As a result, $H=\sqrt{V_{\text{tot}}/3M_P^2}$ will be slightly larger as well. As long as $V\ll V_I$, we can linearize:
\eq{
H(\phi) &= H_I + \Delta H(\phi)\\
&\approx H_I+V(\phi)/(6M_P^2H_I)
}
where $H_I=\sqrt{V_I/3M_P^2}$ is the contribution from the inflaton alone. Decomposing the last term of our Fokker-Planck equation in this way,
\eq{
\dot P(\phi,t) &= \frac1{3H}\partial_\phi(V'(\phi)P(\phi,t))+\frac{H^3}{8\pi^2}\partial^2_{\phi\phi}P(\phi,t)
+3H_I P(\phi,t)+\frac{1}{2M_P^2H_I}V(\phi)P(\phi,t)
}
we see that there is a new $\phi$-dependent term due to this backreaction. At this point, we would like to return to $\rho$, but we encounter  a subtle difficulty. We want $\rho$ to be normalized to 1, while remaining proportional to $P$ at a given time. The appropriate definition is
\eq{
P(\phi,t) = e^{3H_{\text{avg}}t}\rho(\phi,t)
}
where $H_{\text{avg}}$ is the average growth rate of all patches, given by
\eq{
H_{\text{avg}}(t) &= \int H(\phi)\rho(\phi,t)\du\phi
}
While $P$ is described by a local differential equation, this integral means that $\rho$ is not! The dynamics of $\rho$ in a small interval of $\phi$ depend on the average growth rate, which in turn depends on what $\rho$ looks like globally. We will compromise slightly and define an unnormalized distribution $\unrho$ by
\eq{
P(\phi,t) = e^{3H_It}\unrho(\phi,t)
}

The full time-evolution of $\unrho$ is given by the Fokker-Planck equation above, with the $3H_I$ term removed:
\eq{
\dot\unrho(\phi,t) &= \frac1{3H_I}\partial_\phi(V'(\phi)\unrho(\phi,t))+\frac{H_I^3}{8\pi^2}\partial^2_{\phi\phi}\unrho(\phi,t)+\frac{1}{2M_P^2H_I}V(\phi)\unrho(\phi,t)
}

The last term in this equation is due to the backreaction effect.

Note that $\unrho$ is not a probability distribution: the integral $\int\unrho\du\theta$ will change over time at a rate $H_{\text{avg}}-H_I$, which is just the average of $\Delta H$. This normalizes away the growth due to the inflaton but leaves the extra contribution from backreaction. $\unrho$ is useful because it reduces to $\rho$ when backreaction is negligible, while still having local dynamics in $\phi$.

We now substitute $\unrho(\phi,t)=\Psi(\phi)\psi(\phi,t)$, where
\eq{\Psi(\phi)&:=\exp(-\nu(\phi))\\
\nu(\phi)&:=\frac{4\pi^2}{3H_I^4}V(\phi)}
in order to rewrite this diffusion equation as follows, in terms of $\psi$:
\eq{-\frac{4\pi^2}{H_I^3}\dot\psi(\phi,t)=-\frac12\psi''(\phi,t)+\frac12\left[-\nu''(\phi)+\nu'(\phi)^2-\frac{3}{M_P^2}\nu(\phi)\right]\psi(\phi,t)}
This is a Wick-rotated time-dependent Schr\"odinger equation. It has eigenfunction solutions which decay exponentially,
\eq{
\psi(\phi,t) &= \sum_i c_i e^{-\Gamma_i t}\psi_i(\phi)
}
The eigenvalues $\Gamma_i$ and eigenfunctions $\psi_i$ are given by the corresponding time-independent equation
\eq{\frac{4\pi^2}{H_I^3}\Gamma_i\psi_i(\phi)&=-\frac12\psi_i''(\phi)+\frac12\left[-\nu''(\phi)+\nu'(\phi)^2-\frac{3}{M_P^2}\nu(\phi)\right]\psi_i(\phi)}

So far, everything we have said applies to a generic scalar field. For a quadratic potential $V=\frac12m^2\phi^2$, this equation becomes
\eq{
\nu(\phi)&=\frac{2\pi^2m^2}{3H_I^4}\phi^2\\
\frac{4\pi^2}{H_I^3}\Gamma_i\psi_i(\phi)&=-\frac12\psi_i''(\phi)+\frac{\pi^2m^2}{3H_I^4}\left[-2+
\left(\frac{8\pi^2m^2}{3H_I^4}-\frac3{M_P^2}\right)\phi^2\right]\psi_i(\phi)}
For $H_I^4\ll M_P^2m^2$, the backreaction term $3/M_P^2$ is negligible. However, if $H_I^4>\frac{8\pi^2}9M_P^2m^2$, the sign of the ``potential'' flips over, creating an instability. Physically, this means that the patches where the field is further up its potential expand faster by a large enough $\Delta H$ to outpace the relaxation process, so the distribution of patches runs away to extreme field values. If nothing intervenes, this will continue until $\Delta H \approx H_I$, at which point the field $\phi$ becomes a second inflaton field.

For an axion potential, we substitute $\phi=f_a\theta$ and find
\eq{
\nu(\theta)&=\frac{4\pi^2\chi}{3H_I^4}(1-\cos\theta)\\
\frac{4\pi^2f_a^2}{H_I^3}\Gamma_i\psi_i(\theta) &=-\frac12\psi_i''(\theta)+\frac12\left[-\alpha(1-\beta)\cos\theta+\alpha^2\sin^2\theta-\alpha\beta\right]\psi_i(\theta)
}
where
\eq{
\alpha&:=\frac{4\pi^2\chi}{3H_I^4}=\frac12\left(\frac{H_I}{170\MeV}\right)^{-4}\\
\beta&:=\frac{3f_a^2}{M_P^2}
}

The eigenfunctions $\psi_i$ correspond to quasinormal modes given by $\unrho_i=\Psi\psi_i$. The eigenvalues $\Gamma_i$ of these functions are not energies, but decay rates. With the backreaction term neglected, we always have $\Gamma_0=0$ and $\psi_0=\Psi$; the distribution $\unrho_0=\Psi\psi_0=\Psi^2$ is the equilibrium state of the diffusion equation. With backreaction, $\unrho_0$ still gives the behavior at late time, but $\Gamma_0<0$ so the total population grows over time. (This fact is dependent on our choice of zero for $V(\phi)$. We chose $V(\phi)\ge0$, so the backreaction is always a positive contribution to growth and $\Gamma_0$ is negative. For a different choice, our $\Gamma_i$ would all shift by some constant.) The gap $\Gamma_1-\Gamma_0$ gives the rate for $\unrho_1$ to decay relative to $\unrho_0$. This is the slowest-decaying mode, so $1/(\Gamma_1-\Gamma_0)$ is the relaxation time.

Note that the ``potential'' here is not the axion potential. It has a term proportional to $-V''\propto V\propto\cos\theta$, but also another $\propto\sin^2\theta$ with twice the frequency. In addition to the minimum at $\theta=0$, this produces another minimum at $\theta=\pi$.

Setting aside the backreaction effects momentarily, we can use this formalism to compute more precise relaxation times. For low $H_I$ ($\alpha\gg1$), the $\sin^2$ term dominates and we can approximate the system as a simple harmonic oscillator. This approximation gives $\Gamma_0=m_a^2/6H_I$ and $\Gamma_n=(n+\frac12)m_a^2/3H_I$, with a relaxation time of $3H_I/m_a^2$ as expected. (Of course, we should really have $\Gamma_0=0$; it is easy to check that the true ground state $\psi_0\propto\exp(-\nu)$ has $\Gamma=0$. The harmonic-oscillator approximation gives the correct level spacing but not the correct ground-state energy.)

For high $H_I$, the potential is negligible; the eigenfunctions are $\psi_n(\theta)\propto\cos(n\theta)$. It is easy to see that $\psi_0$ is constant, $\Gamma_0=0$, and $\Gamma_1=H_I^3/8\pi^2f_a^2$. This gives a relaxation time of $8\pi^2f_a^2/H_I^3$, which agrees with the estimate in the body of the paper up to a factor of 2.

%backreaction
For $\beta\ll1$ ($f_a\ll M_P$), the effect of backreaction is negligible and the $\theta=0$ minimum dominates. In this regime, the ground state $\psi_0$ of the double-well potential is still approximately $\Psi$.
As $\beta$ approaches $1$, the two minima become closer. At $\beta=1$, the cosine term in the Schr\"odinger potential vanishes and they are exactly degenerate. For $\beta>1$, the $\theta=\pi$ minimum is the global minimum.

This crossover has several effects. First, the ground state becomes a mix of the two minima, then shifts to the $\pi$ minimum. At $\beta=1$ ($f_a=M_P/\sqrt3$), the ground state obeys $\psi_0(0)=\psi_0(\pi)$. However, this is not the phenomenologically-relevant transition: the factor of $\Psi$ still ensures that $\unrho_0(0)\gg\unrho_0(\pi)$. The important transition is at $\beta=2$, where the cosine's coefficient $(1-\beta)$ is the negative of its value in the $f_a\ll M_P$ limit. It is easy to see that this is equivalent to sending $\nu\mapsto -\nu$ or $\theta\mapsto\pi-\theta$, so that $\psi_0\propto\exp(\nu(\theta))\propto 1/\Psi$, and $\unrho_0=\Psi\psi_0=$constant. Therefore, the equilibrium distribution at $\beta=2$ ($f_a=M_P\sqrt2/\sqrt3$) is uniform. (A numerical investigation of the ground state for general $\beta$ suggests that, for $\beta$ approaching this value, the ground state is approximately a mixture of $\Psi$ and a uniform distribution, with the proportions of the mixture changing smoothly.) Above $\beta=2$, for low $H_I$ ($\alpha\gtrsim1$), a Gaussian distribution emerges around $\theta=\pi$ once the peak in $\psi_0$ becomes more significant than the valley in $\Psi$.

Second, the splitting between the lowest two states becomes small (but nonzero), so the relaxation time becomes long. (Phenomenologically, this means that for a narrow window around $f_a=M_P/\sqrt3$ our model is sensitive to initial conditions, just as in the ``eternal inflation only'' window.) At $f_a=M_P\sqrt2/\sqrt3$, the relaxation time returns to the value it would have without the effects of backreaction; for higher $f_a$ it continues to drop.

For low $H_I$ (high $\alpha$), this crossover happens extremely quickly, with the splitting becoming extremely small. As $H_I\to0$ it becomes a first-order phase transition, with the relaxation time diverging at $f_a=M_P/\sqrt3$. (At finite $H_I$ there is technically no phase transition, because no order parameter has a nonanalyticity in $f_a$.)

At high $H_I$ (low $\alpha$), this crossover is unimportant, because the potential is negligible anyway and the relaxation time (splitting) is simply the time for a state to spread across its domain. However, for $\beta\gtrsim1/\alpha$ ($f_a\gtrsim M_P (H_I^2/\Lambda_{QCD}^2)$), the backreaction creates a strong enough potential to concentrate $\unrho$ around $\theta=\pi$; this is transPlanckian enough to be practically irrelevant. The low-$H_I$ and high-$H_I$ behaviors both agree parametrically with the simple estimate of the backreaction transition given in the body of the paper.

\section{Classical Rolling Constraint}
\label{clroll}
To prevent eternal inflation, we can impose the constraint that classical rolling dominates fluctuations for the inflaton itself, which leads to an upper bound on the length of inflation. The fluctuations are of order $H(t)$ per e-fold, so we take the classical-rolling constraint to be, parametrically, $\dot\phi\gtrsim H^2$. More precisely,\cite{Creminelli:2008es}
\eq{
\dot\phi&> \sqrt{\frac3{2\pi^2}}H^2
}
With slow-roll, we obtain a lower bound on $-\dot V$:
\eq{
\dot\phi &\approx \frac{-V'}{3H}\\
-\dot V &= -\dot\phi V'\\
&\approx 3H\dot\phi^2\\
&> \frac9{2\pi^2}H^5}

We can turn this into a lower bound on $-\dot H$:
\eq{
-\dot H &= \frac{-\dot V}{6HM_P^2}\\
&> \frac{3H^4}{4\pi^2M_P^2}}
Note that this gives us a lower bound on the slow-roll parameter $\epsilon$,
\eq{\epsilon &= -\dot H/H^2\\
&> \frac{3H^2}{4\pi^2M_P^2}}
We can also invert it to obtain an upper bound on $N$:
\eq{
-\frac{dt}{dH} &< \frac{4\pi^2M_P^2}{3H^4}\\
-\frac{dN}{dH} &< \frac{4\pi^2M_P^2}{3H^3}\\
N &< \int_{H_f}^{H_i} \frac{4\pi^2M_P^2}{3H^3}dH\\
&=\frac{2\pi^2M_P^2}{3H_f^2}-\frac{2\pi^2M_P^2}{3H_i^2}\\
&< \frac{2\pi^2M_P^2}{3H_f^2}
}
where $H_f$ is the value of $H$ when slow-roll inflation ends and $H_i>H_f$ is the value when it begins.

This bound agrees with the much more general result \cite{Dubovsky:2011uy}, which gives (in three spatial dimensions) that there is a bound on the classical number of e-folds
\eq{
N_c &< S_{dS}/12\\
&= \frac{2\pi^2M_P^2}{3H_f^2}
}
where $S_{dS}$ is the de Sitter entropy at the end of inflation. For larger $N_c$, the reheating volume will be infinite with probability $>0$; in other words, inflation is eternal, at least in some part of the wavefunction for the universe.

Both of these calculations are in terms of $N_c$, which is determined from the purely classical evolution of $\phi$. When using a volume-based measure, the average value of $N$ will be somewhat larger, as the trajectories that fluctuate to higher $H$ will grow faster. (This is analogous to the backreaction effect discussed above, and is the mechanism responsible for eternal inflation when the bound is violated.) Therefore, this bound is actually somewhat pessimistic.

\bibliography{stochastic}

\end{document}